\documentstyle[prl,aps]{revtex}
\begin{document}
\draft
\title{
{}From the defocusing nonlinear Schroedinger to the complex
Ginzburg-Landau equation
}
\author{ Olaf Stiller, Stefan Popp, and Lorenz Kramer}
\address{
Physikalisches Institut der Universit\"at Bayreuth,
D-95440 Bayreuth, Germany
}
\date{\today}
\maketitle

\begin{abstract}
Perturbation approaches developed so far for the dark soliton
solutions of the (fully integrable) defocusing nonlinear Schroedinger equation
cannot describe the dynamics resulting from dissipative
perturbations of the Ginzburg-Landau type.
Here spatially slowly decaying changes of the background wavenumber occur
which requires the use of matching technics.
It is shown how the perturbation selects
a 1 or 2-parameter subfamily from the 3-parameter
family of dark solitons of the nonlinear Schroedinger equation.
The dynamics of the perturbed system can then be described
analytically as motion within this selected subfamily yielding
interesting scenarios. Interaction with shocks occurring in the
complex Ginzburg-Landau equation can be included
in a straight forward way.
\end{abstract}


Submitted to Physica D

\section{Introduction}

The defocusing non-linear Schroedinger equation (NLSE)
\begin{eqnarray}
\partial_{t}  A =i  \left(  \partial_{ x}^2
- | A|^2   \right)  A
\label{NLS}
\end{eqnarray}
 has a 3-parameter
family of dark solitons of the form ${\cal A}={\cal A}_0(x-vt)\exp(-i\Omega
t)$.
In contrast to the pulse-like bright solitons of the focusing case dark
solitons
have a  hole-like shape. In their
center they have a strong depression of the modulus
which decays into plane-wave states for $\zeta:=x-vt \to \pm \infty$.
For arbitrary but fixed $\Omega > 0$ and $Q$ in a range such that
$ \kappa := \sqrt{(\Omega -3Q^2)/2}$ is positive
one can derive systematically the two-parameter
 family of standing soliton solutions
($v=0$)
\begin{eqnarray}
\label{sol}
 {\cal A}_{Q , \Omega}(x,t) & =  & \sqrt{2}
 \left[ \kappa \tanh( \kappa  x) - iQ\right] \exp(i Q
 x - i \Omega t)
\; .
\end{eqnarray}
{}From  Galilean invariance of the NLSE one
then
obtains the
3-parameter family of  soliton solutions
 moving with constant velocity $ v$ ($ \zeta =  x -
v t$):
\begin{eqnarray}
\label{gal}
 {\cal A}_{k, v, \Omega}
( \zeta,t)
&=&
 {\cal A}_{
Q
,
\bar \Omega
 }
( \zeta)
\exp\left[ i(\frac{ v}{2} \zeta
+\frac{ v^2}{4}t)\right]
\end{eqnarray}
where $\bar \Omega = \Omega +\frac{v^2}{4}$.
They
 have the asymptotic  wave number
$ k =  Q + \frac{ v}{2}$.

The aim of this work is to describe the dynamics resulting from a dissipative
perturbation of Ginzburg-Landau type largely as a motion within the three
dimensional soliton space spanned by ($ k, v ,\Omega $).
Thus we take the family parameters as  time-dependent (slow) variables
while other degrees of freedom follow adiabatically.
The problem here is that $-$in contrast to pulses
(where this approach is well known\cite{bright})$-$
 dark solitons do not decay
for large $|\zeta|$ and the wavenumber $k$ of the  asymptotic plane-wave states
cannot be altered globally in
an infinite system.
 Consequently, to our knowledge, only perturbation methods for situations
with  unchanging $k$ have been
developed so far
(see e.g. \cite{direct} and references therein).
This restriction is, however, not applicable  to describe the
phenomena presented  below for
Ginzburg-Landau type perturbations.
 Simulations show  that $k = k (t)$ becomes time dependent
in a \underline {finite}  region around the soliton center (inner
region). In this region
(of size $|(\partial_t k) \zeta |\ll 1$,   see eq.(26) of \cite{new})
 the system remains for all times close to one of the  ($k, v,
\Omega$)-solitons.
The corresponding  \underline {global} solution has to be constructed
by asymptotic matching.
Here, however, in order to describe the
approximate dynamics of the soliton
center
we can restrict our calculations to the inner region
without  performing  the
matching procedure explicitly.
In order to ensure that a global solution exists we will make use of the result
formulated in \cite{new} (section II.C)  that any local solution which does not
diverge
exponentially for $\kappa |\zeta|\gg 1$ (outer region)
can be matched to a global one.
This provides the boundary conditions for $A$ in the inner region
which for small values of $\partial_t k$ overlaps with the outer region.
In particular the procedure  allows to deal with slowly (algebraically)
diverging
terms which arise in the inner region from the change of the wavenumber $k$.

Adding perturbations of Ginzburg-Landau type to the NLSE one obtains
a complex Ginzburg-Landau equation (CGLE)
\begin{eqnarray}
\label{CGLE}
\partial_{ t}  A =
i
\LARGE(
 \partial_{ x}^2  - | A|^2
\LARGE)  A \; + \LARGE( r +\beta \partial_{ x}^2 -\gamma  | A|^2  +\delta   |
A|^4\LARGE) A
\end{eqnarray}
where we wish to consider $r , \beta , \gamma > 0$ and $|\delta | \ll  \gamma$.
It has three parameters that cannot be scaled away,
e.g. $\beta, \gamma$ and $\delta$.
Then $r$ depends on the choice of scale\cite{footnote}.
In general $\delta$ may be complex, but for simplicity we here  assume it to be
real.
The scenarios we are going to describe occur in a parameter regime
where $\gamma \sim O(\beta^2)$. However, since terms containing
$\beta^2$ may not contribute to the calculations due to
PT-symmetry (parity and time reversal) of the NLSE for constantly moving
solutions
(only odd powers of the perturbation
are allowed)
it will be consistent to treat $\gamma$ together with $\beta $
in a first-order perturbation calculus.

Although we assume  $|\delta | \ll \gamma$, which is indeed relevant for
many systems,
 $\delta$ may play a crucial role for the
\underline {qualitative} dynamics of the equation
as was shown in \cite{PRL}.
For $\delta \equiv 0$ a 1-parameter subfamily,
the so-called  Nozaki-Bekki (N.B.) holes,
is
selected
 from the dark solitons.
The  analytical form of this subfamily was presented
first  by Nozaki and Bekki\cite{NB}
(see e.g. section I.A of \cite{new} for the explicit form of these solutions).
They were found to be stable in a belt-like region in parameter space which
is limited by the instability of the asymptotic plane-wave states on one side
and the instability of the core on the other side.
In the NLS limit   the core of the standing hole ($v=0$)
 is stable for  $\gamma < (4/3)\beta^2$
as can be seen from the
 calculations presented below.

For $\delta \neq 0$, however, this subfamily is destroyed everywhere in
parameter space
leaving only the standing ($v=0$) symmetric solution.
 More precisely
simulations show that a N.B. hole which would be stable for $\delta = 0$
is either accelerated ($\delta < 0$) or decelerated ($\delta > 0$).
So for $\delta > 0$ the standing hole persists while for $\delta < 0$ it is
unstable
against
acceleration. $\delta =0$ is thus the threshold of a stationary bifurcation.
In the limit of small velocities the scenario can be described by the
phenomenological equation
\begin{eqnarray}
\label{a}
\dot{v} = \lambda_\delta  \delta v
\end{eqnarray}
where we have used $v $ to parametrize the 1-dimensional subspace of  ($k, v,
\Omega$)
in which the dynamics takes place.
(The dot refers to the time derivative.)
One issue of this work is to determine the phenomenological constant
$\lambda_\delta$
 analytically.

The scenario becomes richer when (with $\delta > 0$) we cross the line in
parameter space
where the core of the standing hole loses its stability.
Then the system undergoes a Hopf bifurcation and the center of the holes start
to oscillate
in space and time.
To describe this oscillatory behavior  one needs two coupled modes.
 As normal form of the
bifurcation one obtains
for  small velocities and small $u$
     \begin{eqnarray}   \label{phen}
\begin{array}{ccccc}
\dot{u} &=& (\lambda -  g v^2)u &+& \delta_1 v     \\
\dot{v} &=&  \mu                u &+& \delta_2 v
\end{array}
     \end{eqnarray}
where $u$ refers to  another (1-parameter) subspace of
($k, v, \Omega$) which will be determined.
The other symbols in eq.(\ref{phen}) refer to constant coefficients
with $\delta_1$ and $\delta_2$
of order $\delta$.
For large negative $\lambda$ ($|\lambda | \gg |\delta |$)
eq.(\ref{a}) is recovered with $\lambda_\delta = \delta_2 -\mu \delta_1 /
\lambda$.
With the perturbation method
presented in the following it is in principle possible to
determine all these phenomenological coefficients quantitatively.
For simplicity we will restrict ourselves to the linear part
 of eqs.(\ref{phen}).

\section{Destruction of the Nozaki-Bekki family}

We start with an analytic proof that the 3-parameter soliton family is indeed
destroyed if $\delta \neq 0$.
{}From the asymptotic states (large $|\zeta|$) alone one already
finds that not more than a 1-parameter subfamily can possibly survive the
Ginzburg-Landau type perturbations.
The selection criteria for this subfamily are\\
i)
The CGLE has a dispersion  relation $\omega =\omega(q)=
r/\gamma +  (1-\beta/\gamma) q^2 + \delta /\gamma
(r/\gamma -\beta/\gamma  q^2 )^2 + O(\delta^2)$
 for plane waves with wave number $q$.
(The frequency
$\Omega$ in the co-moving frame
 is related to the frequency
$\omega$ in the laboratory frame
 via $\Omega=\Omega(q,v) = \omega(q) - v q$.)
Thus $\Omega$ is completely determined by the asymptotic soliton wave number
$k$
and the velocity.
\\
ii) Phase conservation becomes non trivial and relates the asymptotic wave
numbers to the
velocity.\\
To understand this second point one has to realize that the perturbation alters
the symmetry
of the solitons in
a way that the asymptotic wave number $q_1$ for $\zeta\ll 1$ is no longer
equal to that for  $\zeta\gg 1$
 ( $q_2$).
Then phase conservation requires that both sides rotate with the same frequency
$\Omega =
\omega(q_i) - v q_i$ ($i=1,2$) in the co-moving frame.
This yields
\begin{eqnarray}
v=\frac{\omega(q_1) - \omega(q_2)}{q_1 -q_2} = 2(1-\frac{\beta}{\gamma})k
- 2 \delta \frac{\beta}{\gamma^2} (2 \frac{r}{\gamma}
- \frac{\beta}{\gamma}(q_1^2+q_2^2)) k
+ O(\delta^2)
\label{top}
\end{eqnarray}
where we defined $k:= \frac{1}{2}(q_1 +q_2)$.
Below we will use $k$ ($k\propto v + O(\dot k)$) to label the selected soliton
family ${\cal A}_s [k]$.

To see whether the remaining  1-parameter subspace
actually survives we consider  the CGLE eq.(\ref{CGLE}) in the frame moving
with
velocity $v$ ($\zeta=x-vt$) and rotating with frequency $\Omega$.
 For later convenience we multiply the equation by $(\beta - i)$.
Separating the resulting equation into an unperturbed part $F[A]$ and the
perturbation
$I[A]-i\partial_t A + O(\beta^2)$ one obtains by neglecting terms of order
$\beta^2$
\begin{eqnarray}
\label{F0}
 F[ A]&:= &(\partial_{ \zeta}^2
       + i v \partial_{ \zeta}
+ \Omega -|A|^2) A
\\ \nonumber
&=&
i
\left[ (r-\beta\Omega) - (\gamma -\beta)|A|^2 + \delta |A|^4
+ i \beta  v \partial_{ \zeta} -
\partial_t\right] A
:=
I[ A]  - i\partial_t A
\end{eqnarray}
where we assumed $\partial_t A$ to be of order $\beta^1$ in the co-moving
frame.
(As noted above terms containing even powers of the perturbation
do not contribute to our calculations due to symmetry.
So the next relevant corrections to eq.(\ref{F0}) are actually of
order $\beta^3$.)
Now we make the ansatz
\begin{eqnarray}
\label{ansatz}
A(\zeta,t)={\cal A}
\,(\zeta,t) + W(\zeta)
\end{eqnarray}
where  -- in the spirit of the matching approach  --    $ W(\zeta)$ is of order
$\beta^1$
in a finite region
around the hole center.
${\cal A}$ refers to one of the soliton solutions eq.(\ref{gal})  with given
($k,v,\Omega$).
At linear order in $\beta$ one obtains
the inhomogeneous linear differential equation
(valid in the inner region )
\begin{eqnarray}
\label{inhom}
L W = I[
 {\cal A}
]
- i \partial_t
  A
\end{eqnarray}
with
\begin{eqnarray}
 L W
:=
 \frac{\delta F}
{\delta  A}[ {\cal A}]
W
+
 \frac{\delta F}
{\delta { A}^*}[ {\cal A}]
W^*
\end{eqnarray}
for any function $W(\zeta)$.
(The time derivative $ \partial_t A$ in eq.(\ref{inhom})
 -- which has to be zero in the stationary case
considered here first --  will be important in the  next subsections.)
Note that $L$ is selfadjoint.
{}From its neutral modes $i{\cal A}$ and $\partial_\zeta {\cal A}$, which can
be derived from
 gauge  (or 'phase') and  translational invariance of the NLSE, respectively
one can construct a  \underline{localized} mode
$\Psi_{loc}=\partial_\zeta {\cal A} - k i{\cal A}$,  which
decays exponentially for large $|\zeta|$.
Fredholms alternative then yields the solvability condition
for eq.(\ref{inhom}) (with boundary condition that  $W$ should not  diverge
exponentially
in space)
\begin{eqnarray}
\label{cond}
\int_{-\infty}^{+\infty}d\zeta
\,Re\left(( I[
 {\cal A}
 ]
- i \partial_t
  A)
\Psi_{loc}^*\right) = 0
\end{eqnarray}
which contains no boundary terms.
So in principle the destruction of the hole family by the
higher-order perturbation can be demonstrated
 immediately by inserting
the solutions (\ref{gal}) (combined with eq.(\ref{top}) and the dispersion
relation)
into this equation.
In  Appendix \ref{App_A} we derive the identity
\begin{eqnarray}
\label{integrand}
Re\left( I[
 {\cal A}
 ]
\Psi_{loc}^*\right)=
- \delta
 k
\frac{\beta}{\gamma}
( {\cal R}^2 -  {\cal R}_\infty^2)
( {\cal R}^4 -  {\cal R}_\infty^4)
-
\frac{\gamma}{2}
\Delta  v (  {\cal R}_\infty^2 -  {\cal R}^2)^2
=
 \delta k\,
\frac{\beta}{\gamma}({\cal R}_\infty^2-{\cal R}^2)^3
+ O(\delta^2)
\end{eqnarray}
(with ${\cal R} := |
{\cal A}
|$ and
${\cal R}_\infty:={\cal R}( \zeta=\infty)=\Omega + v k - k^2 $)
where
\begin{eqnarray}
\label{delta_v}
\Delta  v := v - 2(1-\beta /\gamma)k\sim O(\delta)
\end{eqnarray}
(compare eq.(\ref{top})) has been defined for later convenience.
To derive the rhs  of eq.(\ref{integrand})
we have made use of the phase-conservation equation
(\ref{top}) which yields in the NLS limit
$\Delta v = -4 \delta k (\beta/\gamma^2) {\cal R}_\infty + O(\delta^2)$.
In the next subsection where perturbations of the
asymptotic  plane-wave states will be taken into account
explicitly this expression will have to be generalized.
Since ${\cal R}^2 < {\cal R}_\infty^2 $ for finite $\zeta$
one immediately sees
from the rhs of eq.(\ref{integrand})
that the solvability condition,
 eq.(\ref{cond}) with $\partial_t A=0$,
is identical with the condition $\delta = 0$
(unless $ k =0$ and thus $ v = 0$).

\section{Acceleration instability}

In the last section we saw that
the perturbed cubic CGLE ($\delta\neq 0$) has no quasi  stationary
solutions with $v\neq 0$ in the vicinity of
the defocusing NLS equation.
The next step now is to make an ansatz for a weakly
time-dependent solution with a free parameter
that can be determined  by
condition
(\ref{cond}).
At values of $\beta$ and $\gamma$ where
(for $\delta= 0$) the  selected subfamily
(or, more precisely, the solutions
presented by Nozaki and Bekki)
is stable we observed numerically that for $0\neq |\delta |\ll \gamma$ the
system moves
in the vicinity of this 1-parameter family
 ${\cal A}_s={\cal A}_s[k]$ with  $\Delta v = O(\delta)$  and
$\Omega=\Omega(k)\left[ + O(\delta )\right]$.
We thus  take the family parameter
$k$ (alternatively one could use $v
\propto k + O(\dot k) $
as family parameter)
as slow variable, assuming that the other degrees of freedom follow
adiabatically.
This leads to the ansatz:
\begin{eqnarray}
\label{At}
\partial_t A = \dot k \partial_k {\cal A}_s[k(t)]
+ O(\delta\beta^3,\delta^2)
 = i\dot { k}
( \zeta {\cal A}_s -
\frac{\beta}{\gamma} \sqrt{2})
+O(\delta\beta^3,\delta^2,v^2)
\end{eqnarray}
(where we used $Q= \frac{\beta}{\gamma} k - \frac{\Delta v}{2} + O(\delta)$)
valid for a finite region around the soliton center (inner region).
For simplicity we
restrict our calculations to the limit of
small velocities (i.e. small $k$) which is of major interest.
It is straight forward (though lengthy)
to generalize them for arbitrary
velocities.

Using eq.(\ref{At}) the integrand
of eq.(\ref{cond})
obtains a supplementary term
proportional to $ \dot { k}$.
One has
\begin{eqnarray}
\label{condacc}
Re\left((I[ {\cal A}_s]- i \partial_t  A)
\Psi_{loc}^*\right)
=
\delta k \frac{\beta}{\gamma}( {\cal R}^2 -  {\cal R}_\infty^2)
 ( {\cal R}^4 -  {\cal R}_\infty^4)
-
\frac{\gamma}{2}
\Delta  v (  {\cal R}_\infty^2 -  {\cal R}^2)^2
+\dot { k}
\left(\frac{ \zeta}{2}\partial_{ \zeta}
 {\cal R}^2
-\frac{\beta}{\gamma}\sqrt{2}
Re(\Psi_{loc})
\right)
\,\, .
\end{eqnarray}
Performing the integration then yields
\begin{eqnarray}
\label{projcond}
- \delta \frac{8}{5}
\frac{\beta}{\gamma} {\cal R}_\infty^4
\,k
-
\frac{2}{3}
\gamma
 {\cal R}_\infty^2
\, \Delta  v
+
(1-2\frac{\beta}{\gamma})
\dot { k}
= 0
\,\, .
\end{eqnarray}
Here $\Delta  v$ has to be determined such that phase
conservation is insured.
While in the time-independent case phase conservation was
given by eq.(\ref{top}) here this equation
has to be altered
by including terms with $\dot { k}$ resulting from the
time dependence (eq.(\ref{At})).
In
Appendix \ref{App_B}
we
derive a generalization of equation
(\ref{top})
valid for arbitrary $\beta$ and $\gamma$.
In the NLS limit it yields
\begin{eqnarray}
\label{top_nls}
\Delta  v = -4\delta {\cal R}_\infty^2\frac{\beta}{\gamma^2} \, k
+\left[
-
h_s
+\frac{\beta}{\gamma^2
}
-\frac{\sqrt{2}{\cal R}_\infty}{ q_0}\frac{\beta}{\gamma}\right]
\frac{\dot {k}}{{\cal R}_\infty^2}
+ O(v^2)
\\
\nonumber
\mbox{with}\quad
h_s:=\frac{{\cal R}_\infty^2 - 2\frac{\beta}{\gamma}  q_0^2}
{2( \beta -\gamma)  q_0^2}
\simeq \frac{9}{4(\beta -\gamma)^3}
-\frac{\beta}{\gamma (\beta -\gamma)}
\end{eqnarray}
where $q_0 := q_2 (v=0) = - q_1 (v=0)$ is the asymptotic
wave number of the perturbed standing soliton.
In the NLS limit
one has
$q_0 = (\sqrt{2}/3) (\gamma - \beta) {\cal R}_\infty+ O(\beta^2
,\delta)$
as can be obtained either from the Nozaki-Bekki solutions,
or, as we show in Appendix \ref{App_B},  by integrating eq.(\ref{proja}).
Inserting  eq.(\ref{top_nls}) into eq.(\ref{projcond})
yields
\begin{eqnarray}
\label{lambda_del}
 \lambda_\delta:= \lim_{ k \to 0}
\frac{\dot{ k}}{ k  \delta}
 = \frac{16}{15}\left[\frac{8}{3} - \frac{2}{3}
\frac{\gamma^2}{\beta} h_s+\frac{2\gamma}{\beta -\gamma}
-\frac{\gamma}{\beta }\right]^{-1}
\Omega^2
\end{eqnarray}
where we inserted ${\cal R}_\infty^2=\Omega$ which can be obtained from
eq.(\ref{gal})
for the standing hole (with $\Delta v=0$).
At $\gamma = (4/3)\beta^2 (1+ O(\gamma/\beta))$
the expression for
$\lambda_\delta$ has a pole. (Including terms of $ O(\gamma/\beta)$
 -- which are neglected for simplicity only --
 actually improves the accuracy of the results,
since terms with $\beta^2$ may not occur.)
As was explained in section III.A of \cite{new}
the pole of $\lambda_\delta$
 indicates the position of the core-instability threshold
so that in the limit $\gamma \sim \beta^2 \to 0$ (for $\delta=0$)
 the core of the standing hole is unstable if
$\gamma > (4/3)\beta^2 (1+ O(\gamma/\beta))$ .
Literally speaking here the hole is accelerated even for $\delta=0$.

\section{Hopf bifurcation}
In the last section we described the acceleration
caused by the quintic term ($\delta\neq 0$)
as a motion within the
one-parameter subfamily
of the three-parameter  family of dark solitons which is selected
under a perturbation of Ginzburg-Landau type.
This description loses its validity as one
approaches the threshold of the core instability.
At the pole of $\lambda_\delta$ (eq.(\ref{lambda_del}))
the prefactor of $\dot { k} $ vanishes
so that one has to include higher-order terms (which contain
$\ddot { k} $ and $\Delta \dot { v}$) in order to balance
the contributions proportional to $\delta $.
This leads to the two-mode scenario described by eqs.(\ref{phen}).

We thus extend eq.(\ref{At}) for the time dependence
by writing
\begin{eqnarray}
\label{At2}
\partial_t A & = & \left(\dot { k} \partial_{ k}
+
\Delta \dot{ v}
\partial_{\Delta  v}
\right)
{\cal A}[k, \Delta v =0,\Omega(k)]
+ \ddot{ k }\partial_{\dot{ k}} W_1
+ O(\delta \beta^2,\delta^2)
\\
\nonumber
&=&
 i\dot { k}
( \zeta {\cal A} -
\frac{\beta}{\gamma} \sqrt{2})
+ i \frac{\Delta \dot{ v}}{\sqrt{2}}
+ \ddot{ k }\partial_{\dot{ k}} W_1
+ O(\delta \beta^2,v^2)
\end{eqnarray}
where $W_1$ is the solution of the equation
\begin{eqnarray}
\label{inhom2}
LW_1 = I_0[
 {\cal A}_s
]
-i\dot k \partial_k
 {\cal A}_s
\end{eqnarray}
($I_0[\,]$ refers to $I[\,]$ with $\delta=0$),
which can be solved for arbitrary $\dot k$  at threshold
of the core instability (i.e. at the pole of $\lambda_\delta$).

At first sight  one would assume
that explicit knowledge of the  perturbation $W_1$ is required
in order to exploit the ansatz (\ref{At2}).
However, to calculate the  phase-conservation condition related to
eq.(\ref{At2})
 one only needs the asymptotic form of
$W_1$
(for $\kappa|\zeta|\gg 1$ )
which
has already
been calculated (Appendix \ref{App_B}) when deriving
eq.(\ref{top_nls}).
Continuing the expansion which led to
eq.(\ref{top_nls} up to the next order
one obtains (see Appendix \ref{App_B})
\begin{eqnarray}
\label{top_osc2}
\Delta  v = -4\delta {\cal R}_\infty^2\frac{\beta}{\gamma^2} \, k
+\left[
-
h_s
+\frac{\beta}{\gamma^2
}
-\frac{\sqrt{2}{\cal R}_\infty}{ q_0}\frac{\beta}{\gamma}\right]
\frac{\dot {k}}{{\cal R}_\infty^2}
+
\frac{\sqrt{2}{\cal R}_\infty}{ q_0}
\Delta  \dot{ v}
-
\left(\frac{2}{\gamma^{2}} + \frac{9}{8(\gamma - \beta)^4}
\right)
\left(\frac{\beta}{\gamma}-\gamma
h_s
\right)
\frac{\ddot {k}}{{\cal R}_\infty^4}\; ,
\end{eqnarray}
which differs from eq.(\ref{top_nls}) by the two last terms on the rhs.
(Now, however, an explicit  result can be given only for the NLS limit since
the second mode labeled by $\Delta  v$ is only known in this
limiting case.)

In order to exploit the solvability condition
(\ref{cond})  we only need to know the integral
over $Re[\partial_{\dot{ k}} W_1
*\Psi_{loc}^*]$.
Symmetry considerations show
 that it
 is an odd function of $ \zeta$
 so that the integral vanishes.
To see this  we consider the symmetry operator
$\Pi[Z(\zeta)]=Z(-\zeta)^*$
(for every complex function $Z(\zeta)$).
Due to PT-symmetry  $F[\,]$, and thus $L$, commute, whereas
 $I[\,]$ anticommutes with  $\Pi[\,]$.
Soliton solutions are eigenfunctions of $\Pi[\,]$
with eigenvalues $+1$ while the neutral modes
of $L$ have the eigenvalue $-1$.
{}From eq.(\ref{inhom2}) one finds that $W_1$ must have eigenvalue $-1$
and  $Re[\partial_{\dot{ k}} W_1 *\Psi_{loc}^*]$ is indeed
an odd function.
So  there appears no contribution containing $\ddot { k}$ in
the  solvability condition eq.(\ref{cond}) and thus eq.(\ref{projcond})
has to be extended only by a term proportional to
$\Delta \dot{ v}
$.
Inserting eq.(\ref{At2}) into  eq.(\ref{cond})
one finds
\begin{eqnarray}
\label{projosc1}
- \delta \frac{8}{5}
\frac{\beta}{\gamma} {\cal R}_\infty^4
\,k
-
\frac{2}{3}
\gamma
 {\cal R}_\infty^2
\, \Delta  v
+
(1-2\frac{\beta}{\gamma})
\dot { k}
+\Delta  \dot{ v}
= 0
\,\, .
\end{eqnarray}
Differentiating equation (\ref{projosc1})
with respect to time yields
\begin{eqnarray}
\ddot {  k}
=
\frac{2 \gamma^2}{3 (\gamma -2\beta)}
{\cal R}_\infty^2
\Delta \dot {  v}
+o(\delta)
\end{eqnarray}
(where $o(\delta)$ contains terms containing $\partial_t^3 k$ and
$\delta\partial_t k$
which are of  order higher than $\delta^1$). This can be used to eliminate
$\ddot { k}$ from eq.(\ref{top_osc2}).
Solving
eqs.(\ref{top_osc2})
and
(\ref{projosc1})
 for $\dot{ k}$ and $\Delta \dot { v}$
one then obtains a set of equations
which can be identified with eqs.(\ref{phen})
by writing $v \propto
  k +
O(\dot k)$ and $u\propto \Delta  v$.
It is thus straight forward
 to extract the somewhat lengthy expressions for  the (linear) coefficients
of eqs.(\ref{phen}) from eqs.(\ref{projosc1}) and (\ref{top_osc2}).
(Nonlinear coefficients can be obtained in principle by the
same method. Calculation, however, are increased considerably.)
We give the expression for the frequency
$\Omega_{osc}=\sqrt{\delta_1\gamma-(\lambda-\delta_2)^2/4}
\simeq \sqrt{\delta_1\gamma} $ of the harmonic oscillations
which eqs.(\ref{phen}) describe  in the limit $0 < \lambda\ll \delta^{0.5}$.
{}From eqs.(\ref{projosc1}) and (\ref{top_osc2}) one obtains:
\begin{eqnarray}
\label{om_osc}
\sqrt{\delta_1\gamma} &=&
\frac{\sqrt{\left[ \frac{8}{5} \frac{\beta}{\gamma}
h_2
- 4\frac{\beta}{\gamma^2}
(1-2\frac{\beta}{\gamma})\right]
\left( 1 - \frac{2}{3} \gamma
h_1\right)
}}{h_2-(1-2\frac{\beta}{\gamma})h_1}
\Omega
\sqrt{{\cal R}_\infty^2 \delta}
\\
\mbox{with }
\qquad
h_1&=&
 \frac{3}{\gamma -\beta}-
\frac{2\gamma^2}{3(\gamma-2\beta)}
\left(\frac{2}{\gamma^{2}} + \frac{9}{8(  \beta -\gamma)^4}
\right)
\left(\frac{\beta}{\gamma}-\gamma
h_s
\right)
\\
h_2&=&\left[ - h_s +\frac{\beta}{\gamma^2}
-
 \frac{3}{\gamma -\beta}
\frac{\beta}{\gamma}\right]
\end{eqnarray}
Using the scale $r=\gamma$ this yields  $\omega_{osc}/\sqrt{\delta} \simeq 2,58
$
for $\beta =0.123$ and $\gamma= 1.66*10^{-2}$.
{}From  simulations of the CGLE we found  $\omega_{osc}/\sqrt{\delta} = 2.5$
$\pm 0.01$ (numerical error)
yielding reasonable agreement in view of nonlinear effects.
For small $\beta$ and $\gamma$ simulations become increasingly time consuming
since the interaction distance between holes and boundaries or sinks increases
(c.f. next section)
so that one needs much longer systems in order to avoid
boundary effects.

\section{Interaction of holes and shocks}

In systems   which contain
more than one hole neighboring holes are generally separated
by shocks. While holes which are sources
(group velocity and thus causality points outward)
and therefore
determine largely the asymptotic wave states
shocks are sinks and thus behave in a rather passive way.
In simulations shocks moving with constant velocities
are formed
when propagating waves with different wave numbers collide.
The velocity of a shock is then determined by the incoming  wave numbers
via eq.(\ref{top}), which  holds for any localized object in the CGLE.

In the presence of a shock the above calculations have to be altered.
If the shock is not too close to the hole
one has a plane-wave solution plus some small perturbation $W$
in the region between the two localized objects.
Getting closer to the shock the perturbation grows and forms the shock.
In order to include the interaction with a shock in the calculations
one has to require that $W$ (as introduced in eq.(\ref{ansatz}))
matches correctly with the shock solution.
While up to now, in order to insure global boundedness,
 we only accepted algebraic growth of $W$  in the inner region,
the matching with the shock requires exponential growth.
The explicit form of $W$ in the region between hole and shock
has to be extracted from the asymptotic form of the shock solution.
In general this requires numerical computation since the shocks
are not known analytically.
If the wave numbers on both sides
of the shock are small the  shock solution can be approximated
by solving the lowest-order phase equation.
In the NLS limit the wave numbers selected by the standing hole
(and for $\gamma \sim \beta^2 \ll 1$ also by the
traveling holes)  are of order $\beta$, and the corresponding
shock solutions can be obtained analytically.
Here we consider the case where a standing hole is perturbed by
a standing shock. The corresponding phase equation for the shock region
reads
\begin{eqnarray}
\partial_t \phi = -\frac{r}{\gamma}
+\gamma^{-1}(1+\beta\gamma) \partial_{xx} \phi
+ (\frac{\beta}{\gamma}-1) (\partial_x \phi)^2
\end{eqnarray}
 which can be solved via a Hopf-Cole transformation.
Looking for constantly moving solutions, i.e. $\partial_t \phi=-\Omega
=-\omega(q_0)$ where $q_0$ is the wavenumber of the standing hole
(see eq.(\ref{q0})),
one obtains for the phase gradient
\begin{eqnarray}
\partial_x \phi = q_0 \tanh\left[\frac{p}{2}(x-L)\right]
\end{eqnarray}
with $p=2(\beta -\gamma)/(1+\beta\gamma) q_0$.
For $x-L \ll -1$ this yields
\begin{eqnarray}
\partial_x \phi \simeq q_0 \left[ 1 - 2 \exp(-|p|L)e^{|p|x}\right]
\end{eqnarray}
Here one sees in particular that $W\sim e^{|p|x}$ grows very slowly
($p\sim\beta^2$)
compared to the fast decay of the  localized mode $\Psi_{loc}$.
Consequently no boundary terms appear in the solvability condition
(\ref{cond}) and eq.(\ref{projcond}) remains valid in the presence of
a shock.
However, since the local wave number is perturbed, the phase-conservation
condition eq.(\ref{top_nls}) is altered.
In a linear theory the change $\Delta v^s$ of
$\Delta v$ which results from the presence of the shock can be superposed
with the contributions proportional to $\delta k$ and $\dot k$
(in the following we denote them by $\Delta v^0$).
This yields
\begin{eqnarray}
\Delta v \left[:= v - (1-\frac{\beta}{\gamma})(q_1+q_2)\right]
= \Delta v^0 - (1-\frac{\beta}{\gamma})q_2^s
\end{eqnarray}
where $q_2^s := q_2-q_0 = -2 q_0 \exp(-|p|L)$
is the change of $q_2$ caused by the shock.
$ \Delta v^0 $ is given by the rhs of  eq.(\ref{top_nls}).
Inserting this
into eq.(\ref{projcond}) one obtains
\begin{eqnarray}
\dot v  &=  &\lambda_\delta \left[ \delta v + \frac{5}{4} \beta^{-1}{\cal
R_\infty}^{-2}
(\beta -\gamma)^2
q_2^s\right] + O(v,\delta v^3,\delta^2,\beta^3)
\\\nonumber
&\simeq &  \lambda_\delta \left[ \delta v +
\frac{5\sqrt{2}}{6\beta }{\cal R_\infty}^{-2}
(\beta-\gamma)^3
e^{-|p|L} \right]
\end{eqnarray}
where $\lambda_\delta $ is given by the rhs of eq.(\ref{lambda_del}).
In particular one finds that the hole-shock interaction is always attractive
(positive acceleration ; $L>0$ has been assumed) which
is consistent with the results (mostly numerical) presented before
(see sections III.D and IV.B of \cite{new}).

\section{Concluding Remarks}
The above analysis was restricted to the parameter range
where the N.B. holes, i.e. the 1-parameter soliton subfamily
which survives in the cubic CGLE, is stable or weakly
unstable.
Here, starting with one of the holes, the system remains
for all times close to the 1-parameter subfamily.
Thus, in the appropriate parameter range, the present work
describes the slowest degree of freedom, which governs the
longtime behavior.
The problem how an arbitrarily initialized soliton  relaxes into this
subfamily has not been considered.
We also dit not treat the destruction of a N.B. hole (into a plane-wave state)
far beyond its stability limit.
In these cases $\Delta v$ does not remain small for
all times, i.e. the system is temporarily far away from the
N.B. family. Then the relevant time scale should be of order
$\beta$ (or $\gamma$) and it is not necessary to include higher-order
perturbations to the cubic CGLE.
We believe, however, that the method which combines a solvability
condition for the core region with a phase-conservation condition
obtained from the far field, should in principal be applicable
also to more general situations.

We wish to thank Igor Aranson for enlightning discussions.
Support by the Deutsche Forschungsgemeinschaft
(Kr-690/4, Schwerpunkt
'Strukturbildung in dissipativen kontinuierlichen Systemen:
Experiment und Theorie im quantitativen Vergleich')
is gratefully acknowledged.
\begin{appendix}
\section{destruction of the family}
\label{App_A}

In the following we derive eq.(\ref{integrand}).
For this purpose
we multiply
eq.(\ref{F0}) (with $\partial_t A = 0$ ) successively
by $-i A^*$ and $\psi_{loc}^*
:=\partial_{ \zeta} A^* + ik A^*$.
Defining the "current" $j_{loc}:= Re(i A\partial_{ \zeta} \psi_{loc}^*)$
(this yields $j_{loc}
= (\partial_{ \zeta} \phi -k)R^2$ if $R( \zeta):= | A( \zeta)|\neq 0$
; $\phi$ refers to the  Phase of $A$)
one can write for the real parts of the resulting equations:
\begin{eqnarray}
\label{proja}
\partial_{ \zeta} (j_{loc} +  Q  R^2) =
\hat f(R^2)R^2
- \beta  v j_{loc}
\end{eqnarray}
\begin{eqnarray}
\label{projb}
\partial_{ \zeta}\left(
| \psi_{loc} |^2 - \int^{ R^2} f(u)du
\right)
&=&
\hat f(R^2) j_{loc}
 - \beta v | \psi_{loc} |^2
\equiv
Re(
I[A]\psi_{loc}^*
)
\; .
\end{eqnarray}
Here $f:= \omega - k^2  -  R^2$ and $\hat f:=r  -\beta\omega -(\gamma -\beta)
R^2
+ \delta   R^4$ (with $\omega = \Omega + v k$) are polynominals of $R^2$.
For any soliton solution ${\cal A}$ which survives the perturbation
eq.(\ref{cond}) (with $\partial_t A = 0$) states that the integral over the rhs
of eq.(\ref{projb}) has to be zero if
$ A =  {\cal A}$ (then we write
$\Psi_{loc}, J_{loc}, {\cal R} $
for
$\psi_{loc}, j_{loc}, R) $.

In the unperturbed case
(i.e. $\beta =\gamma = \delta = r=0$ so that the rhs of eqs.(\ref{proja},
\ref{projb})
vanish)
one finds from eq.(\ref{proja}) that  $ J_{loc} -   Q {\cal R}^2$
is a constant.
Since $ J_{loc}$  vanishes for $\zeta\rightarrow \pm\infty$
by construction  we can write
\begin{eqnarray}
\label{J}
J_{loc} =   Q( {\cal R}^2 -  {\cal R}_\infty^2)
\; .
\end{eqnarray}
Similarly one finds from eq.(\ref{projb}) that $|\Psi_{loc}|^2$ is a second
order polynomial of
${\cal R}^2$.
The double zero for
$\zeta=\pm \infty$
allows us to write
\begin{eqnarray}
\label{Bx}
|\Psi_{loc}|^2 =
\frac{1}{2}
( {\cal R}_\infty^2 -  {\cal R}^2)^2
\,\, .
\end{eqnarray}
{}From eq.(\ref{proja}) one can
further
 see that also
$\hat f( {\cal R}^2)$ has to vanish asymptotically, i.e.
\begin{eqnarray}
\label{hatf}
\hat f( {\cal R}^2) =
(\gamma -\beta)(  {\cal R}_\infty^2 -  {\cal R}^2)
+  \delta ( {\cal R}^4 -  {\cal R}_\infty^4)
\end{eqnarray}
in order to avoid divergencies of $j_{loc}$.
Combining eqs.(\ref{J}), (\ref{hatf}) and (\ref{Bx})
we can write the rhs of eq.(\ref{projb}) at lowest
order in $\beta$
\begin{eqnarray}
\hat f( {\cal R}^2)J_{loc} - \beta  v | {\cal A}_{ \zeta} |^2
&= &
 \delta
  Q
( {\cal R}^2 -  {\cal R}_\infty^2)
( {\cal R}^4 -  {\cal R}_\infty^4)
-
\frac{\gamma}{2}\Delta  v (  {\cal R}_\infty^2 -  {\cal R}^2)^2
\end{eqnarray}
where the rhs can be identified with the lhs of
 eq.(\ref{integrand}) by noting that  $ Q = \frac{\beta}{\gamma}   k
+ O(\delta)$ holds, as can be seen from eq.(\ref{top}) and the definition
$ Q =  k -\frac{ v}{2}$ (see eq.(\ref{gal})).

\section{Phase conservation in the time dependent case}
\label{App_B}

Here we show how eq.(\ref{top})
has to be altered when the asymptotic plane-wave states
are perturbed.
We do this by solving
eq.(\ref{CGLE}) with the ansatz  (\ref{At})
  explicitly in the overlap region
where $\kappa |\zeta| \gg 1$ (so that hole solutions can be
approximated by a plane wave) but
$|(\partial_t k) \zeta |\ll 1$
(so that eq.(\ref{At}) for the time dependence is still valid).
Analogously to the derivation of eq.(\ref{top})
we will require that the solution derived for the
regions of overlap holds on both sides independently.
 As long as both sides are
antisymmetric
 (like for
the standing hole) phase conservation
is fulfilled trivially.
When the symmetry of the standing hole is destroyed
phase conservation becomes non trivial and one obtains
 a relation between the velocity and the symmetry breaking
parts of the solution (i.e. $k$ and $\dot k$, compare eqs.
 (\ref{top}) and (\ref{top_nls}) ).

The following calculation
is not restricted to the NLS limit
if one generalizes the ansatz (\ref{At}) by writing
\begin{eqnarray}
\label{AtNB}
\partial_t A = \dot k \partial_k {\cal A}^{NB}[k(t)]
+ O(\delta^2)
\end{eqnarray}
where ${\cal A}^{NB}$ refers to the
Nozaki and Bekki solutions \cite{NB}
(see e.g. Appendix A of \cite{new}).
We start from eq.(\ref{CGLE}) in the moving and constantly rotating frame
and separate into real  and imaginary part
using $A= R \exp[i\phi]$
\begin{eqnarray}
\label{eR}
\begin{array}{ccccccccccccc}
\dot { R} &=&
(r-\beta\phi_\zeta^2 -\gamma R^2 + \delta R^4) &R&
-&  &[\phi_{\zeta\zeta} R +2\phi_\zeta R_\zeta]&
+ & v& R_\zeta &+&\beta &R_{\zeta\zeta}
\\
\label{ep}
\dot { \phi}  R &=&
(\Omega -\phi_\zeta^2 - R^2 ) &R&
+&\beta  &[\phi_{\zeta\zeta} R +2\phi_\zeta R_\zeta]&
+ & v &\phi_\zeta R  &+& &R_{\zeta\zeta}
\end{array}
\end{eqnarray}
For $\kappa |\zeta|\gg 1$
the Nozaki-Bekki  solutions can be approximated by a plane-wave solution
(of the CGLE with $\delta=0$)
with some wave number $q$ ($q=q_{2/1}$ for $\pm \kappa \zeta \gg 1$).
One has
\begin{eqnarray}
\label{pw}
R^2 &=& \rho_0^2(q) := \frac{1}{\gamma}(r-\beta q^2)
\\
\nonumber
\phi &=& q \zeta
+ \varphi
\end{eqnarray}
where $\varphi $ has to be extracted from the Nozaki-Bekki solutions.
In the NLS limit one has
$\varphi=\mp \arcsin\sqrt{2}\rho_0^{-1} (\frac{\beta}{\gamma}k-\frac{\Delta v
}{2})$ for the
left/right overlap region which can be obtained from the soliton solutions
(\ref{gal}).
Consistently with eq.(\ref{At}) the time dependence of $R$ and $\phi$ at
lowest order is given by
\begin{eqnarray}
\label{Rt}
\dot { R} &=&
-
\frac{\beta}{\gamma}
\frac{q}{\rho_0}
\dot k
\;
+ O(v^2)
\\
\label{pt}
\dot { \phi}  &=&
(\zeta+ \partial_k\varphi)
\dot k
\;
+ O(v^2)
\end{eqnarray}
which corresponds to a change of the wave number ($\dot q=\dot k + O(v^3)$)
plus a change
of the phase difference.
Since eq.(\ref{pt}) contains a diverging term the solutions
 $  R$ and $ \phi$ of eqs.(\ref{eR})
also have to diverge in the overlap region.
We thus make the following ansatz:
\begin{eqnarray}
\label{aR}
 R &=&  \rho +  \rho_l  \zeta
\\
\label{ap}
\partial_\zeta  \phi &=&  q +  q_l  \zeta
\end{eqnarray}
where $ \rho_l ,    q_l $ vanish for $\dot { k}\rightarrow 0$
  (we note that $\dot { k} \sim \delta [v + O(v^3)] $).
Inserting this into eqs.(\ref{eR})
one obtains linear terms $\sim  \zeta^1$ and constant terms  $\sim
\zeta^0$whose coefficients have to cancel respectively.
Terms proportional to $ \zeta^2$ can be neglected since they are of
order higher than $\dot { k}$.

{}From the linear terms one obtains at order $\delta^1$:
\begin{eqnarray}
\label{lx1}
&
\begin{array}{ccccc}
0 &=& -2\beta q q_l &-& 2\gamma \rho_0\rho_l
\\
\label{lx2}
\dot { k}  &=&  -2q q_l &-& 2\rho_0\rho_l
\end{array}
&
\\
\mbox{or}
\qquad
&
\label{rol}
 \rho_l=\rho_l(q;\dot k) =
\frac{\dot k}{2(\frac{\gamma}{\beta}-1)\rho_0}
\label{ql}
\; ,
\qquad
  q_l
= q_l(q;\dot k) =
\frac{\dot k}{2(\frac
{\beta}
{\gamma}
-1)q}
&
\,\, .
\end{eqnarray}

Constant terms yield
\begin{eqnarray}
\label{ro}
 \rho^2
&=&\rho^2(q;\dot k) = \frac{1}{\gamma}
\left( r -\beta  q^2 + \delta \rho_0^4
- \left[q_l + 2 \frac{q \rho_l}{\rho_0}\right]
+ \frac{\beta}{\gamma}\rho_0^{-2} q \dot k
\right)
+ O(v^2)
\\
\mbox{and}
\qquad
\label{v1}
 v  q &=&
g_{\dot k}(q):=
-\omega +  q^2 +  \rho^2
-\beta
\left[q_l + 2 \frac{q \rho_l}{\rho_0}\right]
+
\dot k
 \partial_k\varphi
+ O(v^2)
\,\, .
\end{eqnarray}
In analogy with eq.(\ref{top}) we now write
\begin{eqnarray}
\label{v3}
 v = \frac{g_{\dot k}(q_1) + g_{\dot k}(q_2)}
{ q_1 -  q_2}
\end{eqnarray}
which yields
\begin{eqnarray}
\label{top_nls_a}
\Delta  v = -4\delta\frac{\beta}{\gamma^2}\rho_0^2\, k
+\left[
- (1+\beta\gamma )h_s
+\frac{\beta}{\gamma^2\rho^2}
+\frac{\partial_k\varphi}{q_0}
\right]\dot {k}
+ O(v^2)
\\
\nonumber
\mbox{with}\quad
h_s=\frac{\rho_0^2 - 2\frac{\beta}{\gamma}  q_0^2}
{2(\gamma -\beta)  q_0^2\rho_0^2}
\end{eqnarray}
where
 $q_0:= q_2 (v=0) = - q_1 (v=0)$
refers to the wave number of the standing Nozaki-Bekki solution
which can be obtained either from these solutions or  -- in the NLS limit --
also by integrating eq.(\ref{proja}) (with $k=v=0, \delta=0$) at order
$\beta^1$
\begin{eqnarray}
\label{q0}
q_0\simeq\frac{\left. j_{loc}\right|_{-\infty}^{+\infty}}{2{\cal R}_{\infty}^2}
\simeq\frac{1}{2{\cal R}_{\infty}^2}\int_{-\infty}^{+\infty} \hat f({\cal R}^2)
{\cal R}^2 d\zeta =
\frac{\sqrt{2}}{3} (\gamma -\beta ){\cal R}_{\infty}
\end{eqnarray}
where terms of order $\beta^2$ have been neglected.

\vspace{0.5cm}

At the next order the time dependence is given
by eq.(\ref{At2}). For the overlap region this
yields
\begin{eqnarray}
\label{Rt2}
\dot { R} &=&
-
\frac{\beta}{\gamma}
\frac{q}{\rho_0}
\dot k
+
\frac{1}{2}\dot \rho_l
\zeta^2
\\
\label{pt2}
\dot { \phi}  &=&
(\zeta+ \partial_k\varphi)
\dot k
+
\frac{1}{2}\dot q_l
\zeta^2
+
\Delta \dot v
\partial_{\Delta v}
\varphi
\end{eqnarray}
where the change of the phase $\partial_{\Delta v} \varphi$
 proportional
to $\Delta v$ is only known in the NLS limit.
{}From eqs.(\ref{rol})
one finds
\begin{eqnarray}
\label{rolt}
\dot \rho_l &=&
\frac{\ddot k}{2(\frac{\gamma}{\beta}-1)\rho_0}
+O(v^2)
\\
\label{qlt}
 \dot q_l &=&
\frac{\ddot k}{2(\frac {\beta} {\gamma} -1)q}
+O(v^2)
\,\, .
\end{eqnarray}
Equations (\ref{eR})
with (\ref{Rt2}) and (\ref{pt2})
can be solved by the ansatz:
\begin{eqnarray}
\label{aR2}
 R &=&  \rho +  (\rho_l+\rho_{l2})  \zeta
+\rho_{ll} \zeta^2
\\
\label{ap2}
\partial_\zeta  \phi &=&  q +(  q_l+q_{l2})  \zeta
+q_{ll} \zeta^2
\end{eqnarray}
where $\rho_{l2}$, $q_{l2}$, $\rho_{ll}$ and
$q_{ll}$ are proportional
 to $\ddot{ k}$.
(terms containing  $\dot{ k}^2$
are neglected since they are of order $O(v^2)$.)
Collecting terms proportional to $\zeta^1$ and $\zeta^2$
yields
at order  $\ddot{ k}$
four linear equations from which the quantities
 $\rho_{l2}$, $q_{l2}$, $\rho_{ll}$ and $q_{ll}$
can be determined.
{}From the constant terms ($\sim \zeta^0$) one  obtains
expressions
for $\rho^2$
and for $vq$ which differ from eqs.(\ref{ro}) and (\ref{v1})
only by terms proportional to  $\ddot{ k}$.
Then the same procedure as in the last section
(see eq.(\ref{v3}))
leads to
eq.(\ref{top_osc2}).

\end{appendix}


\begin{thebibliography}{99}
\bibitem  [\dag] {footnote}    The notation of the previous article \cite{new}
                  is recovered by rescaling: $b=1/\beta$, $c=1/\gamma$
                     and $d=\delta \gamma/r^2$ .
\end{thebibliography}
\end{document}